\documentclass[aps,prb,amssymb,showpacs,superscriptaddress,twocolumn,
tightenlines]{revtex4}%
\usepackage{graphicx}
\usepackage{amsmath,bm}
\usepackage[mathscr]{euscript}
\usepackage{dcolumn}% Align table columns on decimal point

\begin{document}
\title{Role of the Ward Identity and Relevance of the $\bm{G_0W_0}$ 
Approximation\\
in Normal and Superconducting States}

\author{Yasutami Takada}
\thanks{Email: takada@issp.u-tokyo.ac.jp; to be published in Molecular Physics; 
DOI:10.1080/00268976.2015.1131860}
\affiliation{Institute for Solid State Physics, University of Tokyo,
             5-1-5 Kashiwanoha, Kashiwa, Chiba 277-8581, Japan}
%\address{Institute for Solid State Physics, University of Tokyo,
%         5-1-5 Kashiwanoha, Kashiwa, Chiba 277-8581, Japan}

%\date{\today}

%%%%%%%%%%%%%%%%%%%%%%%%%%%< Abstract: 236 words >%%%%%%%%%%%%%%%%%%%%%%%%%%%%%%
\begin{abstract}
On the basis of the self-consistent calculation scheme for the electron 
self-energy $\Sigma$ with the use of the three-point vertex function $\Gamma$ 
always satisfying the Ward identity, we find that the obtained quasiparticle 
dispersion in the normal state in gapped systems such as semiconductors, 
insulators, and molecules is well reproduced by that in the one-shot GW 
(or G$_0$W$_0$) approximation. In calculating the superconducting transition 
temperature $T_c$, we also find a similar situation; the result for $T_c$ in 
the gauge-invariant self-consistent (GISC) framework including the effect of 
$\Gamma$ satisfying the Ward identity is different from that in the conventional 
Eliashberg theory (which amounts to the GW approximation for superconductivity) 
but is close to that in the G$_0$W$_0$ approximation. Those facts indicate that 
the G$_0$W$_0$ approximation actually takes proper account of both vertex and 
high-order self-energy corrections in a mutually cancelling manner and thus we 
can understand that the G$_0$W$_0$ approximation is better than the fully 
self-consistent GW one in obtaining some of physical quantities. 
\end{abstract}

%67.85.LM Degenerate Fermi gases
%71.10.Ca Electron gas, Fermi gas 
%71.15.Mb Density functional theory, 
%	local density approximation, gradient and other corrections 
%71.45.Gm Exchange, correlation, dielectric and magnetic response functions, 
%         plasmons 
%71.55.-i Impurity and defect levels 
%75.20.Hr Local moment in compounds and alloys; Kondo effect, 
%         valence fluctuations, heavy fermions 
%74.70.Tx Heavy-fermion superconductors (for heavy-fermion systems in 
%         magnetically ordered materials, see 75.30.Mb; see also 
%71.27.+a Strongly correlated electron systems, heavy fermions) 
%71.15.Pd Molecular dynamics calculations (Car-Parrinello) and other 
%         numerical simulations 
%74.10.+v Occurrence, potential candidates 
%74.20.-z Theories and models of superconducting state 
%74.20.Fg BCS theory and its development 
%74.20.Mn Nonconventional mechanisms 
%74.20.Pq Electronic structure calculations (for methods of electronic 
%         structure calculations, see 71.15.-m) 
%74.70.Ad Metals; alloys and binary compounds (including A15, MgB2, etc.) 

\maketitle

%\section{}
%\label{}

%%%%%%%%%%%%%%%%%%%%%%%%%%%%%%%%< Introduction >%%%%%%%%%%%%%%%%%%%%%%%%%%%%%%%%
\section{Introduction}

%%%%%%%%%< Paragraph 1: Geneal introduction, with focusing on \Sigma >%%%%%%%%%%
A nonperturbative self-consistent approach to the electron self-energy $\Sigma$ 
was provided in 1965 by Hedin~\cite{Hedin65} in a closed set of equations, 
relating $\Sigma$ with the one-electron Green's function $G$, the dynamic 
screened interaction $W$, the polarization function $\Pi$, and the vertex 
function $\Gamma$. This is a formally exact formulation, but it is difficult 
to implement this scheme as it is, because we cannot determine the electron-hole 
irreducible interaction $\tilde{I}$, a central quantity in the Bethe-Salpeter 
equation to calculate $\Gamma$, through its original definition using a 
functional derivative, $\tilde{I}=\delta \Sigma/\delta G$. Thus we are 
forced to adopt some approximate treatments such as the GW approximation 
in which $\Gamma$ is taken as unity. 

%%%%%%%%%%%%%%%%%%%%%%%%< Paragraph 2: G_0W_0A vs GWA >%%%%%%%%%%%%%%%%%%%%%%%%%
For more than two decades, successful calculations have been done for molecules, 
clusters, semiconductors, and insulators in the one-shot GW (or G$_0$W$_0$) 
approximation~\cite{Hybertsen85,Hybertsen86,Aryasetiawan98,Aulbur00,Ishii,Kikuchi}, 
but this is usually regarded as a too primitive approximation, mostly because 
it is, in general, not a conserving approximation in the sense of Baym and 
Kadanoff~\cite{BK1,BK2}. In contrast, the fully self-consistent GW approximation 
obeys the conservation laws related to the macroscopic quantities like the total 
electron number. In recent years, this self-consistent calculation has become 
feasible, but we are led to a puzzling conclusion that the experiment on 
quasiparticle properties in semiconductors and insulators is much better 
described in the G$_0$W$_0$ approximation than in the GW 
one~\cite{Eguiluz98,Delaney04}. In atoms and molecules, the situation is less 
clear, but in many cases the G$_0$W$_0$ approximation gives better 
results~\cite{Stan09,Rostgaard10,Caruso13,Koval14}. 

\begin{figure}
\begin{center}
\includegraphics[scale=0.44,keepaspectratio]{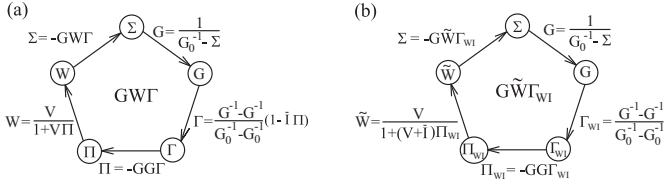}
\end{center}
\caption[Fig.1]{Self-consistent iteration loops to determine the electron 
self-energy (a) in the original GW$\Gamma$ and (b) in its improved version 
or the G$\tilde{\rm W}\Gamma_{WI}$, respectively.}
\label{fig:1}
\end{figure}

%%%%%%%%%%%%%%%%%%%%< Paragraph 3: Metals and vertex part >%%%%%%%%%%%%%%%%%%%%%
In metals, on the other hand, neither G$_0$W$_0$ nor GW works very 
well~\cite{Kutepov09}, requiring us to include $\Gamma$ in some way in treating 
systems possessing gapless excitations. Some schemes have already been proposed 
for this purpose~\cite{Bruneval05,Shinshkin07}, but they do not satisfy the Ward 
identity (WI), an exact relation between $\Sigma$ and $\Gamma$ due to gauge 
invariance representing the local electron-number conservation~\cite{Takahashi57}. 
In 2001, based on general consideration on algorithms beyond the Baym-Kadanoff 
one~\cite{Takada95a}, a scheme was proposed to incorporate $\Gamma$ in the GW 
approximation with automatically fulfilling the WI~\cite{Takada01}. This 
GW$\Gamma$ scheme with the use of the information on the local-field factor in 
the electron gas~\cite{Richardson94} for determining $\bar{I}$ (see, 
Fig.~\ref{fig:1}(a)) succeeded in obtaining the correct quasiparticle behavior 
in simple metals. In 2011, this scheme was further improved~\cite{Maebashi11} 
into the G$\tilde{\rm W}\Gamma_{WI}$ scheme (see, Fig.~\ref{fig:1}(b)) so as to 
avoid the problem of dielectric catastrophe associated with the divergence of 
the compressibility $\kappa$ at the electron density specified by $r_s$ equal 
to 5.25 in the electron gas and concomitantly that of the static $\Pi$ in the 
long wave-length limit~\cite{Takada05,Maebashi09}. This scheme is numerically 
confirmed to provide nonnegative one-electron spectral functions 
$A({\bm p},\omega)$ for the homogeneous electron gas at least for $r_s \leq 8$ 
without any special treatments to impose on the positivity of 
$A({\bm p},\omega)$, contrary to a recently proposed scheme~\cite{Stefanucci14}.

%%%%%%%%%%%%%%%%%%< Paragraph 4: Aim for the normal state >%%%%%%%%%%%%%%%%%%%%%
In the former part of this paper, we study the quasiparticle properties in the 
normal state to find that the quasiparticle dispersion self-consistently obtained 
in the G$\tilde{\rm W}\Gamma_{WI}$ scheme for semiconductors and insulators 
is essentially the same as that in the G$_0$W$_0$ approximation, implying that 
G$_0$W$_0$ is superior to GW in the sense that for the systems with gapful 
excitations, it actually takes proper account of the mutual cancellation 
between vertex and high-order self-energy corrections. This observation resolves 
the above-mentioned long-standing puzzle on GW versus G$_0$W$_0$. Here we 
emphasize that this cancellation is considered up to infinite order as a whole 
with specifying the assumptions needed in the cancellation, in sharp contrast 
with the claims of a similar kind in the past~\cite{DuBois1,DuBois2,Takada91}; 
they were inferred from the behavior of low-order terms in perturbation 
expansion for metals. 

%%%%%%%%%%%%%%%%%%%%%%%%%%%%%%%%%%< Figure 1 >%%%%%%%%%%%%%%%%%%%%%%%%%%%%%%%%%%
%------------------------------------------------------------------------------%

%%%%%%%%%%%%%%< Paragraph 5: Aim for the superconducting state >%%%%%%%%%%%%%%%%
In the latter part of this paper, we study a similar problem in the 
superconducting state by the calculation of its transition temperature $T_c$ 
in the phonon mechanism in both the conventional Eliashberg 
theory~\cite{Eliashberg60,McMillan68,Allen75,Allen82,Carbotte90} and the 
gauge-invariant self-consistent (GISC) method~\cite{Takada93,Takada95b} which 
correspond, respectively, to the GW and GW$\Gamma$ schemes for calculating the 
normal-state properties. Even for the case of weakly correlated and weakly 
coupled superconductors such as Al, we find that the calculated $T_c$ in GISC is 
different from that in the Eliashberg theory but close to that in the G$_0$W$_0$ 
approximation. We discuss the implication of this result from various aspects, 
including the importance of determining the Coulomb pseudopotential 
$\mu^*$~\cite{Morel62} from first principles and a proposal of the suitable 
functional form for the pairing interaction kernel in the density functional 
theory for superconductors (SCDFT)~\cite{Oliveira88,Kurth99,Luders05,Marques05}.

%%%%%%%%%%%%%%%%%%%%%%%%%%%%%%%%< Normal State >%%%%%%%%%%%%%%%%%%%%%%%%%%%%%%%%
\section{Normal state}

%%%%%%%%%%%%%%< Paragraph 6: Formulation: Exact set of equations >%%%%%%%%%%%%%%
For systems with translation symmetry in which momentum ${\bm p}$ is a good 
quantum number, the exact Hedin's relations can be explicitly written in the 
following way: The Dyson equation relates $G(p)$ with $\Sigma(p)$ through 
$G(p)^{-1} \!=\! {G_0(p)}^{-1}\!-\!\Sigma(p)$ with $p$ a combined notation of 
${\bm p}$, spin $\sigma$, and fermion Matsubara frequency $i\omega_p \! \equiv 
\! i\pi T (2p\!+\!1)$ at temperature $T$ with an integer $p$~\cite{units}. 
The bare Green's function $G_0(p)$ is written as $G_0(p)\!=\!(i\omega_p\!-\!
\varepsilon_{{\bm p}})^{-1}$ with $\varepsilon_{{\bm p}}$ the bare one-electron 
dispersion. The Bethe-Salpeter equation determines $\Gamma(p\!+\!q,p)$ by 
\begin{align}
\label{eq:1}
\Gamma(p\!+\!q,p)= 1 + \sum_{p'} &\tilde{I}(p\!+\!q,p;p'\!+\!q,p')
\nonumber \\
&\times G(p')G(p'\!+\!q)\Gamma(p'\!+\!q,p'), 
\end{align}
where $q$ is a combined notation of ${\bm q}$ and boson Matsubara frequency 
$i\omega_q \! \equiv \! i2\pi Tq$ with $q$ an integer and $\sum_{p'}$ 
represents the sum $T\sum_{\omega_{p'}}\! \sum_{\bm p'}\sum_{\sigma'}$. 
By using $\Gamma(p\!+\!q,p)$, we can give $\Pi(q)$ and $\Sigma(p)$ by
\begin{align}
\label{eq:2}
\Pi(q) &= - \sum_{p}G(p\!+\!q)G(p)\Gamma(p\!+\!q,p),
\\
\label{eq:3}
\Sigma(p) &=-\sum_{q}G(p\!+\!q)W(q)\Gamma(p\!+\!q,p),
\end{align}
respectively, with $\sum_{q}=T\sum_{\omega_{q}}\! \sum_{{\bm q}}$ and 
$W(q)=V({\bm q})/[1+V({\bm q})\Pi(q)]$, where $V({\bm q})$ is the bare 
Coulomb interaction $4\pi e^2/{\bm q}^2$. 

%%%%%%%%%%< Paragraph 7: Formulation: Average irreducible interaction >%%%%%%%%%
In Ref.~\cite{Takada01}, the concept of ``the ratio function'' was introduced 
to obtain a good {\it approximate} functional form for $\Gamma(p\!+\!q,p)$ 
satisfying the WI. By exploiting this concept, we have explored an {\it exact} 
functional form for $\Gamma(p\!+\!q,p)$ and succeeded in obtaining the following 
form: 
\begin{eqnarray}
\label{eq:4a}
\Gamma(p\!+\!q,p)=\Gamma^{(a)}(p\!+\!q,p)\Gamma^{(b)}(p\!+\!q,p),
\end{eqnarray}
where $\Gamma^{(a)}(p\!+\!q,p)$ and $\Gamma^{(b)}(p\!+\!q,p)$ are, respectively, 
defined as 
\begin{align}
\label{eq:5a}
\Gamma^{(a)}(p\!+\!q,p)&\equiv 1-\langle \tilde{I} \rangle_{p+q,p}\Pi(q),
\\
\label{eq:5}
\Gamma^{(b)}(p\!+\!q,p)  &\equiv  
\frac{{G(p\!+\!q)}^{-1} \! - \! {G(p)}^{-1}} 
{{G_0(p\!+\!q)}^{-1}  \!- \! {G_0(p)}^{-1} \!- \!\Delta\Sigma_{p\!+\!q,p}}.
\end{align}
Here an average of $\tilde{I}$, $\langle \tilde{I} \rangle_{p+q,p}$, 
and a difference in the self-energy, $\Delta\Sigma_{p\!+\!q,p}$, are, 
respectively, introduced by
\begin{align}
\label{eq:4}
\langle \tilde{I} \rangle_{p+q,p} & \equiv -\sum_{p'} 
\tilde{I}(p\!+\!q,p;p'\!+\!q,p')
\nonumber \\
& \hskip 1.2truecm \times G(p')G(p'\!+\!q)\Gamma(p'\!+\!q,p')/\Pi(q),
\\
\label{eq:6}
\Delta\Sigma_{p\!+\!q,p}  & \equiv 
\sum_{p'}\!\tilde{I}(p\!+\!q,p;p'\!+\!q,p')[G(p')\! - \! G(p'\!+\!q)],
\end{align}
as functionals of $G$ and $\tilde{I}$. If $\tilde{I}$ is exact, 
$\Gamma^{(a)}(p\!+\!q,p)$ is nothing but $\Gamma(p\!+\!q,p)$ in Eq.~(\ref{eq:1}), 
as can easily be seen from the very definition of $\langle \tilde{I} 
\rangle_{p+q,p}$, and $\Delta\Sigma_{p\!+\!q,p}$ is reduced to 
$\Sigma(p\!+\!q)\!-\!\Sigma(p)$, leading to $\Gamma^{(b)}(p\!+\!q,p)\!=\!1$. 
Thus Eq.~(\ref{eq:4a}) provides the same $\Gamma(p\!+\!q,p)$ as that in the 
Hedin's exact theory. In reality, the exact $\tilde{I}$ is not known and we have 
to employ some approximate $\tilde{I}$, in which an advantage of Eq.~(\ref{eq:4a}) 
over Eq.~(\ref{eq:1}) becomes apparent; the former provides $\Gamma(p\!+\!q,p)$ 
satisfying the WI irrespective of the choice of $\tilde{I}$, while the latter 
does not. 

%%%%%%%%%%%%%%< Paragraph 8: Formulation: Reduction to GW\Gamma >%%%%%%%%%%%%%%%
Physically $\tilde{I}$ takes care of exchange and correlation effects in 
$\Gamma(p\!+\!q,p)$ and it is well known that this physics can be captured 
by the local-field factor for the homogeneous electron gas or by the Jastrow 
factor for inhomogeneous systems. In either way, these effects are well described 
in terms of a function depending only on the inter-electron distance, which 
justifies to assume that $\tilde{I}(p\!+\!q,p;p'\!+\!q,p')$ depends only on $q$ 
to write $\tilde{I}(p\!+\!q,p;p'\!+\!q,p')=\bar{I}(q)$. If this assumption is 
adopted in our exact framework, we obtain $\langle \tilde{I} \rangle_{p+q,p}=
\bar{I}(q)$ and $\Delta\Sigma_{p\!+\!q,p}=0$. Then, by defining 
$\Gamma_{WI}(p\!+\!q,p)$ by 
\begin{eqnarray}
\label{eq:7}
\Gamma_{WI}(p\!+\!q,p) \equiv 
\frac{{G(p\!+\!q)}^{-1} \! - {G(p)}^{-1}}
{{G_0(p\!+\!q)}^{-1} - {G_0(p)}^{-1}}, 
\end{eqnarray}
we obtain $\Gamma(p\!+\!q,p)=[1-\bar{I}(q)\Pi(q)]\Gamma_{WI}(p\!+\!q,p)$, a 
result given in Ref.~\cite{Takada01}, leading to the GW$\Gamma$ scheme in 
Fig.~\ref{fig:1}(a). 

%%%%%%%%%%%%%%%%< Paragraph 9: Formulation: Modified GW\Gamma >%%%%%%%%%%%%%%%%%
By putting this form of $\Gamma(p\!+\!q,p)$ into Eq.~(\ref{eq:2}), 
we find that $\Pi(q)$ is written as
\begin{eqnarray}
\label{eq:8}
\Pi(q) = \frac{\Pi_{WI}(q)}{1+\bar{I}(q) \Pi_{WI}(q)},
\end{eqnarray}
with $\Pi_{WI}(q)$, defined by
\begin{eqnarray}
\label{eq:9}
\Pi_{WI}(q) = - \sum_{p}G(p\!+\!q)G(p)\Gamma_{WI}(p\!+\!q,p).
\end{eqnarray}
Then we can rewrite $\Sigma(p)$ in Eq.~(\ref{eq:3}) into 
\begin{eqnarray}
\label{eq:10}
\Sigma(p) =-\sum_{q}G(p\!+\!q)\tilde{W}(q)\Gamma_{WI}(p\!+\!q,p),
\end{eqnarray}
with $\tilde{W}(q) \equiv V({\bm q})/\{1+[V({\bm q})+\bar{I}(q)]\Pi_{WI}(q)\}$. 
Combining these results, we can construct the G$\tilde{\rm W}\Gamma_{WI}$ scheme 
as shown in Fig.~\ref{fig:1}(b). This scheme is equivalent to the GW$\Gamma$ one 
in obtaining $\Sigma(p)$, but computational consts are much reduced by the 
calculation of $\Pi(q)$ through Eq.~(\ref{eq:8}) via $\Pi_{WI}(q)$, because 
Eq.~(\ref{eq:9}) can be cast into a form convenient for numerical calculations as 
\begin{eqnarray}
\label{eq:11}
\Pi_{WI}(q) \equiv \Pi_{WI}({\bm q},i\omega_q) 
= \sum_{{\bm p}\sigma}\frac{n({\bm p\!+\!\bm q})\!-\!n({\bm p})}
{i\omega_q \!-\!\varepsilon_{{\bm p\!+\!\bm q}}\!+\!\varepsilon_{{\bm p}}},
\end{eqnarray}
where $n({\bm p})\,[=T\sum_{\omega_{p}}\!G(p)e^{i\omega_p 0^+}]$ is 
the momentum distribution function in the interacting system. Note that this 
expression very much resembles the one for the polarization function in the 
random-phase approximation (RPA) $\Pi_{0}(q)$, which is given by
\begin{eqnarray}
\label{eq:12}
\Pi_{0}(q) \!=\! -\! \sum_{p}\!G_0(p)G_0(p\!+\!q)
\!=\! \sum_{{\bm p}\sigma}\!\frac{f(\varepsilon_{{\bm p\!+\!\bm q}})\!-\!
f(\varepsilon_{{\bm p}})}
{i\omega_q \!-\!\varepsilon_{{\bm p\!+\!\bm q}}\!+\!\varepsilon_{{\bm p}}},
\end{eqnarray}
where $f(\varepsilon)$ is the Fermi distribution function or the 
momentum distribution function in the non-interacting system. 

%%%%%%%%%%%%%< Paragraph 10: Mention on conserving approximation >%%%%%%%%%%%%%%
With the use of Eq.~(\ref{eq:7}) and introducing ${\tilde \varepsilon}_{\bm p}$ by
\begin{eqnarray}
\label{eq:13a}
{\tilde \varepsilon}_{\bm p}
= \varepsilon_{\bm p}\!-\!\sum_{q} \frac{\tilde{W}(q)}
{i\omega_q \!-\!\varepsilon_{{\bm p\!+\!\bm q}}\!+\!\varepsilon_{{\bm p}}}, 
\end{eqnarray}
we can rewrite our scheme into an integral equation to determine $G(p)$ through 
\begin{eqnarray}
\label{eq:13}
(i \omega_p\!-\!{\tilde \varepsilon}_{\bm p})G(p) = 1\!+\!\sum_{q}
\frac{\tilde{W}(q) G(p\!+\!q)}{i\omega_q \!-\!\varepsilon_{{\bm p\!+\!\bm q}}\!
+\!\varepsilon_{{\bm p}}}\,.
\end{eqnarray}
On the assumption of $\bar{I}(q)=0$, this equation coincides with the one for 
obtaining the asymptotically exact $G(p)$ in a neutral Fermi system such as 
the one-dimensional Tomonaga-Luttinger model~\cite{DL74,Maebashi14} or 
higher-dimensional models with strong forward scatterings~\cite{MCC98}. 
This coincidence clearly demonstrates the intrinsically nonperturbative 
nature of our framework.

%%%%%%%%%%%%%< Paragraph 11: Choice of the local-field factor >%%%%%%%%%%%%%%%%%
In principle, $\bar{I}(q)$ is at our disposal, but Eq.~(\ref{eq:8}) suggests us 
to choose $\bar{I}(q)=-G_{+}(q)V({\bf q})$ with $G_{+}(q)$ the local-field 
factor. Note, however, that the meaning of $G_{+}(q)$ here is different 
from the ordinary one that is defined with respect to $\Pi_{0}(q)$ 
instead of $\Pi_{WI}(q)$. Fortunately, we already know a good form for 
$G_{+}(q)$ with taking account of this difference, which is $G_s(q)$ in 
Ref.~\cite{Richardson94}, satisfying the exact limit due to 
Niklasson~\cite{Niklasson74} as $|{\bm q}| \to \infty$. The self-consistent 
results for the homogeneous electron gas up to $r_s=8$ with this choice of 
$\bar{I}(q)$ are given in Ref.~\cite{Maebashi11}. 

%%%%%%%%%%%%%%%%%%%%%%%< Paragraph 12: Crystalline case >%%%%%%%%%%%%%%%%%%%%%%%
In the crystalline case, each quantity involved in the G$\tilde{\rm W}\Gamma_{WI}$ 
scheme should be represented in the matrix form with respect to the 
reciprocal-lattice vectors $\{{\bm K}\}$. For example, $G(p)$ is a matrix 
composed of the elements $\{ G_{\bm K,\bm K'}({\bm p},i\omega_p) \}$ with 
${\bm p}$ a wave vector in the first Brillouin zone. For some quantities, we 
need to add the conversion factors transforming from the plane-wave basis to 
the Bloch-function one in considering the matrix elements; for example, 
${\Pi_{0\,}}_{\bm K,\bm K'}(q)$ is given as
\begin{align}
\label{eq:14}
&{\Pi_{0\,}}_{\bm K,\bm K'}(q) \!=\! \sum_{nn'{\bm p}\sigma}\!
\frac{f(\varepsilon_{n'{\bm p\!+\!\bm q}})\!-\!f(\varepsilon_{n{\bm p}})}
{i\omega_q \!-\!\varepsilon_{n'{\bm p\!+\!\bm q}}\!+\!\varepsilon_{n{\bm p}}}
\nonumber \\
& \hskip 0.5truecm \times \!
\langle n\!{\bm p}|e^{-i({\bm q\!+\!\bm K})\cdot {\bm r}}|n'\!{\bm p\!+\!\bm q}
\rangle \langle n'\!{\bm p\!+\!q}|e^{i({\bm q\!+\!\bm K'})\cdot {\bm r'}}
|n\!{\bm p}\rangle,
\end{align}
where $|n\!{\bm p}\rangle$ is the Bloch function for the 
$n$th band~\cite{Adler62,Wiser63}. 

%%%%%%%%%%%%< Paragraph 13: Quasiparticle energy in the GW\Gamma >%%%%%%%%%%%%%%
With this understanding, we can apply the G$\tilde{\rm W}\Gamma_{WI}$ scheme to 
semiconductors and insulators possessing a gap in the electronic excitation 
energies. Then, without detailed computations, the self-consistently 
determined quasiparticle energy $E_{{\bm p}}$ in this scheme is found to be 
well approximated by that in the G$_0$W$_0$ approximation, as we shall explain 
in the following. 

%%%%%%%%< Paragraph 14: Quasiparticle dispersion in the GW\Gamma_WI >%%%%%%%%%%%
Let us assume that $\Pi_{WI}(q)=\Pi_0(q)$ and $\bar{I}(q)=0$ for the time being. 
Then, we may rewrite Eq.~(\ref{eq:10}) as 
\begin{eqnarray}
\label{eq:15}
\Sigma(p) = 
-\!\sum_q \!\frac{W_0(q)}{G_0(p\!+\!q)^{-1} \!-\! G_0(p)^{-1}}
+ \gamma (p) G(p)^{-1},
\end{eqnarray}
with $W_0(q) \equiv V({\bm q})/[1+V({\bm q})\Pi_{0}(q)]$ and $\gamma (p)$ 
a dimensionless function, defined by
\begin{eqnarray}
\label{eq:16}
\gamma (p) \! \equiv \! \gamma ({\bm p}, i \omega_p) \!=\! \sum_q 
\frac{G(p+q)W_0(q)}{
i\omega_q \!-\!\varepsilon_{{\bm p\!+\!\bm q}}\!+\!\varepsilon_{{\bm p}}}.
\end{eqnarray}
The quasiparticle dispersion $E_{{\bm p}}$ is determined by the pole of the 
retarded one-electron Green's function $G^R({\bm p},\omega)$, or 
$G^R({\bm p},E_{{\bm p}})^{-1}=0$, amounting to $E_{{\bm p}}=
\varepsilon_{\bm p}+\Sigma^R({\bm p},E_{{\bm p}})$, where we obtain the 
``on-shell'' retarded self-energy as 
\begin{eqnarray}
\label{eq:17}
\Sigma^R({\bm p},E_{{\bm p}}) = -\sum_q \frac{W_0(q)}
{i\omega_q \!-\!\varepsilon_{{\bm p\!+\!\bm q}}\!+\!\varepsilon_{{\bm p}}}, 
\end{eqnarray}
by analytic continuation of $\Sigma(p)$ in Eq.~(\ref{eq:15}). In deriving 
Eq.~(\ref{eq:17}), we have paid due attention to the convergence of 
$\gamma^R ({\bm p},E_{{\bm p}})$ in gapful systems. In fact, provided that 
$\bar{I}(q)=0$, ${\tilde \varepsilon}_{\bm p}$ and the integral in the right-hand 
side in Eq.~(\ref{eq:13}) are, respectively, reduced to $E_{{\bm p}}$ and 
$\gamma(p)$, leading to the behavior of $G^R({\bm p}, \omega)$ for $\omega$ 
near $E_{{\bm p}}$ as 
\begin{eqnarray}
\label{eq:18}
G^R({\bm p}, \omega) \approx 
\frac{1\!+\!\gamma^R ({\bm p}, E_{{\bm p}})}{\omega +i0^+- E_{{\bm p}}}. 
\end{eqnarray}

%%%%%%%%%%%%< Paragraph 15: Quasiparticle energy in the G_0W_0 >%%%%%%%%%%%%%%%%
For comparison, let us consider the self-energy in the G$_0$W$_0$ approximation, 
which is given by $\Sigma_0(p)\!=\!-\!\sum_q \!G_0(p\!+\!q)W_0(q)$. By analytic 
continuation $i \omega_p \to \epsilon_{\bm p}\!+\!i0^+$, we obtain 
\begin{align}
\label{eq:19}
\Sigma_0^R({\bm p},\varepsilon_{\bm p}) \!=&\!
-\!\sum_q \!\frac{W_0(q)}{i\omega_q \!-\!\varepsilon_{{\bm p\!+\!\bm q}}\!
+\!\varepsilon_{\bm p}}
\nonumber \\
& \!-\frac{1}{2}\! \sum_{\bm q}\! 
W_0({\bm q}, \varepsilon_{{\bm p\!+\!\bm q}} \!-\! \varepsilon_{\bm p})
\nonumber \\
& \times \Bigl [ \coth \frac{\varepsilon_{{\bm p\!+\!\bm q}} 
\!-\! \varepsilon_{\bm p}}{2T}\!- 
\tanh \frac{\varepsilon_{{\bm p\!+\!\bm q}}}{2T} \Bigr].
\end{align}
Because the transition ${\bm p\!+\!\bm q}\to {\bm p}$ involved in Eq.~(\ref{eq:19}) 
is relevant only for the interband transition, $|\varepsilon_{{\bm p\!+\!\bm q}} 
\!-\! \varepsilon_{\bm p}|$ is always larger than $E_g$ the energy gap. At low $T$, 
the chemical potential $\mu$ lies at the center of the band gap, indicating that 
$|\varepsilon_{{\bm p\!+\!\bm q}}| \!\ge\! E_g/2$. These two facts allow us to 
safely neglect the contribution from the second sum in Eq.~(\ref{eq:19}), as 
long as $T \ll E_g$. Thus we may write $E_{{\bm p}}^0$ the 
quasiparticle dispersion in the G$_0$W$_0$ approximation as 
\begin{align}
\label{eq:20}
E_{{\bm p}}^0 = \varepsilon_{{\bm p}} + 
\Sigma_0^R({\bm p},\varepsilon_{{\bm p}}) 
=\varepsilon_{{\bm p}} -\sum_q \frac{W_0(q)}
{i\omega_q \!-\!\varepsilon_{{\bm p\!+\!\bm q}}\!+\!\varepsilon_{{\bm p}}},
\end{align}
leading us to conclude that $E_{{\bm p}}^0 = E_{{\bm p}}$. Note, however, that 
the spectral weight $z_{\bm p}\, [=(1\!-\!\partial \Sigma_0^R({\bm p},\omega)/
\partial \omega)^{-1}|_{\omega=\varepsilon_{{\bm p}}}]$ is different from 
$1\!+\!\gamma^R ({\bm p}, E_{{\bm p}})$. 

%%%%%%%%%%%%%%%%%%%%%%%%%<  Paragraph 16: z-factor >%%%%%%%%%%%%%%%%%%%%%%%%%%%%
In the literature, $E_{{\bm p}}^0$ is sometimes evaluated as $E_{{\bm p}}^0=
\varepsilon_{{\bm p}}+z_{\bm p}\Sigma_0^R({\bm p},\varepsilon_{{\bm p}})$ and 
there is a controversy as to whether this $z_{\bm p}$ should be included or not. 
As previously discussed in detail~\cite{Takada91}, we consider it better not 
to include $z_{\bm p}$ so that the vertex corrections beyond the RPA are 
properly included, together with higher-order self-energy terms in a mutually 
cancelling manner. In fact, our present result of $E_{{\bm p}}^0 = E_{{\bm p}}$ 
without this factor $z_{\bm p}$ indicates that this feature of mutual 
cancellation reaches far up to infinite order in semiconductors and insulators. 

%%%%%%%%%%%%%%%%%%%< Paragraph 18a: Comment on \Pi_WI=\Pi_0 >%%%%%%%%%%%%%%%%%%%
Finally we comment on the two assumptions as well as other related issues: 
(i) The difference between $\Pi_{WI}$ and $\Pi_0$ arises only from that between 
$n({\bm p})$ and $f(\varepsilon_{\bm p})$. In usual semiconductors and insulators, 
the valence-electron density is high; for example, $r_s=2$ for Si. Now $n({\bm p})$ 
in a metal at such $r_s$ does not deviate much from $f(\varepsilon_{\bm p})$ 
except for the states near the Fermi level, as shown in Fig. 4 in 
Ref.~\cite{Maebashi11}, but those states are absent from the outset in these 
gapful systems. Thus $n({\bm p})$ is close to $f(\varepsilon_{\bm p})$, leading 
to $\Pi_{WI} \approx \Pi_0$. 
%%%%%%%%%%%%%%%%%%%%< Paragraph 18b: Comment on \bar{I}=0 >%%%%%%%%%%%%%%%%%%%%%
(ii) Justification of $\bar{I}=0$ has already been done by numerical studies 
in Ref.~\cite{Hybertsen86}, in which $\bar{I}$ in our scheme is critically 
assessed in terms of $K_{xc}$ the density-derivative of the Kohn-Sham 
exchange-correlation potential. From an analytic point of view, it is enough 
to note that the basic processes to contribute to $\bar{I}$ are related to the 
interband electron-hole interactions, in which $|{\bm q}|$ for principal 
processes is of the order of $|{\bm K}|$, making $V({\bm q})$ very small and 
$G_{+}(q)$ reach its asymptotic constant. Thus the effect of $\bar{I}$ is weak 
in semiconductors and insulators. 
%%%%%%%%%%%%%%%%%%%%< Paragraph 18c: Large gap system  >%%%%%%%%%%%%%%%%%%%%%%%%
(iii) The inherent problem in the G$_0$W$_0$ approximation is the dependence of 
the results on the arbitrary starting basis. From our perspective, the 
dependence originates from the degree of satisfying the above two assumptions. 
Thus, if the gap energies (or the vertical ionization potentials (IPs) for atoms 
and molecules) are large enough, the dependence becomes small and the 
results in the G$_0$W$_0$ approximation from any starting point are close to 
the experimental values, as seen for IPs of He and Ne in Table V in 
Ref.~\cite{Koval14} and of CO and N$_2$ in Table I in Ref.~\cite{Rostgaard10}.
%%%%%%%%%%%%%%%%%%%%< Paragraph 18d: Large gap system  >%%%%%%%%%%%%%%%%%%%%%%%%
(iv) We have shown that the G$_0$W$_0$ approximation is good for obtaining the 
quasiparticle dispersion for gapful systems, but there is no guarantee 
for other physical quantities such as the line shape of $A({\bm p},\omega)$. 
On the other hand, the framework stipulated by the set of equations, 
Eqs.~(\ref{eq:2})-(\ref{eq:6}), is, in principle, exact and can provide 
accurate results for the physical quantities directly derived from $G(p)$. Thus 
the physical requirements such as the nonnegativity of $A({\bm p},\omega)$ are 
automatically satisfied, as long as we choose an appropriate approximation to 
$\tilde{I}(p\!+\!q,p;p'\!+\!q,p')$ which is a single quantity in the framework 
to control the accuracy of the results. We can determine a proper functional 
form for $\tilde{I}(p\!+\!q,p;p'\!+\!q,p')$ by the use of the information on 
the local-field factor or by perturbation expansion from either weak- or 
strong-coupling limit, supplemented by the information obtained by quantum 
Monte Carlo simulations, if available. 

%%%%%%%%%%%%%%%%%%%%%%%%%%%%< Superconducting State >%%%%%%%%%%%%%%%%%%%%%%%%%%%
\section{Superconducting state}

%%%%%%%%%%%%%    Paragraph 19: Formalism in Nambu representation     %%%%%%%%%%%
In the conventional phonon mechanism of superconductivity with spin-singlet 
$s$-wave Cooper pairing, we can formulate the problem in much the same way as 
in the normal state, if we employ the Nambu representation~\cite{Nambu60}. 
We obtain the rigorous expressions for $\Pi(q)$ and $\Sigma(p)$ as
\begin{align}
\label{eq:21}
\Pi (q) &= -\sum_{p} {\rm Tr} \left [ \tau_3 G(p+q)\Gamma (p+q,p) G(p) \right ],
\\
\label{eq:22}
\Sigma(p) &= -\sum_{q} \tau_3 G(p+q) \Gamma(p+q,p) W(q),
\end{align} 
respectively. They are very similar to Eqs.~(\ref{eq:2}) and (\ref{eq:3}), but 
$\Sigma(p)$, together with the scalar vertex function $\Gamma(p+q,p)$, is now 
a $2 \times 2$ matrix. Here, $\tau_i$s for $i$=1, 2, and 3 are the usual 
$2 \times 2$ Pauli matrices and $W(q)$ is the effective electron-electron 
interaction. For the electron-phonon coupled system, $W(q)$ is exactly obtained as
\begin{align}
W(q) = \frac{V({\bm q})+V_{ph}(q)}{1+[V({\bm q})+V_{ph}(q)]\Pi (q)},
\label{eq:23}
\end{align} 
where $V_{ph}(q)$ is the bare phonon-mediated electron-electron interaction, 
given by
\begin{align}
V_{ph}(q) = \sum_{\nu} {|g_{{\bm q}\nu}|}^2 
\frac{2 \Omega_{{\bm q}\nu}}
{{(i\omega_q)}^2 - {\Omega_{{\bm q}\nu}}^2},
\label{eq:24}
\end{align} 
with bare electron-phonon coupling $g_{{\bm q}\nu}$ and bare phonon energy 
$\Omega_{{\bm q}\nu}$ for the $\nu$th phonon. In this Nambu representation, 
the WI is written in the form of 
\begin{align}
(i\omega_{p'} - i\omega_p)&\Gamma(p',p)-({\bm p'}-{\bm p}) \!\cdot \!
\mbox{\boldmath$\Gamma$}(p',p)
\nonumber \\
&= {G(p')}^{-1} \tau_3 - \tau_3 {G(p)}^{-1}, 
\label{eq:26}
\end{align} 
where $\mbox{\boldmath$\Gamma$}(p',p)$ is the vector vertex function. 
Without loss of generality, the $2\times 2$ matrix $\Sigma(p)$ can be 
decomposed into 
\begin{align}
\Sigma(p) = [1-Z(p)]i\omega_p\tau_0 + \phi (p)\tau_1 + \chi (p)\tau_3, 
\label{eq:27}
\end{align} 
with $\tau_0$ the unit matrix. Then we may rewrite Eq.~(\ref{eq:26}) into 
the form as 
\begin{align}
(i\omega_{p'} \!-\! i\omega_p)&\Gamma(p',p)-({\bm p'}\!-\!{\bm p}) \!\cdot \!
\mbox{\boldmath$\Gamma$}(p',p)
\nonumber \\
&=[i\omega_{p'}Z(p')\!-\!i\omega_p Z(p)]\tau_3 
-[\tilde{\varepsilon}(p')\!-\!\tilde{\varepsilon}(p)]\tau_0
\nonumber \\
&\hskip 0.4truecm +[\phi (p') \!+\! \phi (p)]i \tau_2,
\label{eq:28}
\end{align} 
with $\tilde{\varepsilon}(p)$ defined by $\tilde{\varepsilon}(p) \equiv 
\varepsilon_{\bm p}+\chi (p)$. 

%%%%%%%%%%%%%%%%%%%%%   Paragraph 20: Eliashberg theory  %%%%%%%%%%%%%%%%%%%%%%%
Now let us assume that the average phonon energy $\langle \Omega \rangle$ 
is much smaller than the Fermi energy $\varepsilon_F$ of conduction electrons, 
as is usually the case for superconductors in the phonon mechanism. In this 
situation, it is physically relevant to separate the purely Coulombic part 
from $W(q)$ in Eq.~(\ref{eq:23}) in such a way as
\begin{align}
W(q)=&\frac{V({\bm q})}{1+V({\bm q})\Pi (q)}
+\frac{1}{[1+V({\bm q})\Pi (q)]^2}
\nonumber \\
&\times \frac{V_{ph}(q)}{1+V_{ph}({\bm q})\Pi (q)/[1+V({\bm q})\Pi (q)]}.
\label{eq:23a}
\end{align} 
In Eq.~(\ref{eq:23a}), the second term repersents the fully screened 
phonon-mediated interaction, which is supposed to play a central role in 
bringing about superconductivity. Because this interaction works only in the 
energy range of about $\langle \Omega \rangle$, the purely Coulombic term, 
which extends over the range of $\varepsilon_F$, is usually renormalized and 
reduced into the Coulomb pseudopotential $\mu^*$ which is supposed to work 
only in the range of about $\langle \Omega \rangle$ in the gap equation to 
determine $T_c$~\cite{Morel62}. The actual value for $\mu^*$ will be determined 
phenomenologically in order to reproduce the experimental value of $T_c$. 
Upon these assumptions, $W(q)$ is not zero only in the vicinity of the Fermi 
level. This situation can well be treated by the introduction of the cutoff 
energy $\omega_c$ $(\ll \varepsilon_F)$ in considering the sum over momentum 
${\bm p}$ in such a way as
\begin{align}
\sum_{\bm p}&=\int N(\varepsilon) d\varepsilon,\ \text{with}\ 
N(\varepsilon) \equiv \sum_{{\bm p}} 
\delta (\varepsilon - \varepsilon_{{\bm p}}) 
\nonumber \\
&= 
\begin{cases}
N(0) & \text{for}\ |\varepsilon|<\omega_c, \\
\ 0  & \text{otherwise,}
\end{cases}
\label{eq:25b}
\end{align} 
where $N(\varepsilon)$ is the electronic density of states per one spin. 

\begin{figure}[htbp]
\begin{center}
\includegraphics[scale=0.42,keepaspectratio]{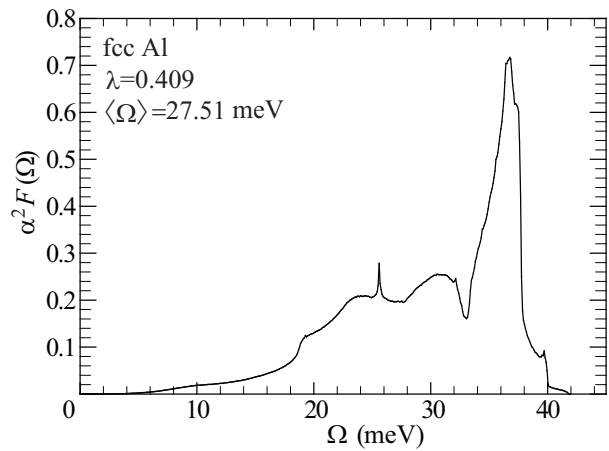}
\end{center}
\caption[Fig.2]{The Eliashberg function for fcc Al.}
\label{fig:2}
\end{figure}

%%%%%%%%%%%%%%%%%%%%%   Paragraph 21: Eliashberg function  %%%%%%%%%%%%%%%%%%%%%
Traditionally, a further simplification is made by assuming that $W(q)$ is 
independent of ${\bm q}$. Then $W(q)=W(i\omega_q)$ is written in terms of 
$\alpha^2 F(\Omega)$ the so-called Eliashberg function as
\begin{align}
& W(i\omega_q) = \frac{\mu^*-\lambda(\omega_q)}{N(0)}
\nonumber \\
&\text{with}\ 
\lambda(\omega_q) \equiv 
\int^{\infty}_0 d\Omega\, \alpha^2 F(\Omega) \,
\frac{2\Omega}{{\omega_q}^2 + {\Omega}^2}.
\label{eq:25}
\end{align} 
Note that $\lambda=\lambda(0)$ is the usual nondimensional electron-phonon 
coupling constant and $\langle \Omega \rangle$ is given by
\begin{align}
\langle \Omega \rangle=\frac{2}{\lambda}
\int^{\infty}_0 d\Omega\, \alpha^2 F(\Omega).
\label{eq:25a}
\end{align} 
The definition of the Eliashberg function and its calculated results 
on various superconductors are available in the literature~\cite{Savrasov96}, 
but in recent years we can obtain the results rather easily by using 
the packages for first-principles calculations such as Quantum Espresso~\cite{QE}.
In Fig.~\ref{fig:2}, we show an example of the calculated $\alpha^2 F(\Omega)$, 
which is obtained for fcc Al with the mesh of $48\times 48 \times 48$ for 
${\bm p}$ and $16\times 16 \times 16$ for ${\bm q}$ in the first Brilouin zone. 

%%%%%%%%%%%%%%%%%%%%%%%%%%%%%%%%%%< Figure 2 >%%%%%%%%%%%%%%%%%%%%%%%%%%%%%%%%%%
%------------------------------------------------------------------------------%

%%%%%%%%%%%%%%%%  Paragraph 22: Momentum-independent situation  %%%%%%%%%%%%%%%%
Incidentally, under the condition of the momentum-independent interaction in 
Eq.~(\ref{eq:25}) and the elecron-hole symmmetric situation as suggested in 
Eq.~(\ref{eq:25b}), the level-shift function $\chi(p)$ in Eq.~(\ref{eq:27}) is 
easily found to be zero. The renormalization function $Z(p)$ and the gap 
function $\phi(p)$ are not zero but independent of the momentum variable 
${\bm p}$. Then we can assume that the scalar vertex function is also 
independent of momentum variables and it must obey the relation, given by
\begin{align}
(i\omega_{p'} \!-\! i\omega_p)&\Gamma(i\omega_{p'},i\omega_p)
=[i\omega_{p'}Z(i\omega_{p'})\!-\!i\omega_p Z(i\omega_p)]\tau_3 
\nonumber \\
&+[\phi (i\omega_{p'}) \!+\! \phi (i\omega_p)]i \tau_2.
\label{eq:40}
\end{align} 
This relation is derived by setting ${\bm p'}={\bm p}$ in Eq.~(\ref{eq:28}) and 
indicates that $\Gamma(i\omega_{p'},i\omega_p)$ contains the $\tau_2$-component. 
As originally discussed by Nambu~\cite{Nambu60}, this component is related to the 
phase-collective Nambu-Goldstone mode. In superconductors the energy of this 
mode is the plasmon energy $\omega_{pl}$, which is the energy scale far beyond 
$\langle \Omega \rangle$. Thus we will omit this contribution to the vertex 
function in the present treatment in which all physical quantities beyond 
the energy scale $\langle \Omega \rangle$ will be neglected with the hope that 
due effects will be renormalized and included into the definition of $\mu^*$. 
There are also discussions on the $\tau_1$-component in the vertex function, 
which is related to the amplitude-collective Nambu-Goldstone 
mode~\cite{LV82,Varma15}, but it is known that this contribution does not 
change $T_c$~\cite{Nambu94}, assuring us of neglecting it altogether in the 
calculation of $T_c$. Thus we obtain $\Gamma(i\omega_{p'},i\omega_p)$ as 
\begin{align}
\Gamma(i\omega_{p'},i\omega_p) 
= \frac{i\omega_{p'} Z(i\omega_{p'}) - i\omega_p Z(i\omega_p)}
{i\omega_{p'} - i\omega_p}\, \tau_3 ,
\label{eq:30}
\end{align} 
for $\omega_{p'} \neq \omega_p$. At $\omega_{p'} = \omega_p$, we cannot use 
the WI to determine it, but on general considerations, it is given by 
$[1-\delta \Sigma(p)/\delta \mu]\tau_3$ with the normal-state self-energy 
$\Sigma(p)$ and $\mu$ the Fermi level, which is different from 
$\lim_{\omega_{p'} \to \omega_p}\Gamma(i\omega_{p'},i\omega_p)$ in 
Eq.~(\ref{eq:30}). In this paper, however, we shall use this limit, i. e., 
$\Gamma(i\omega_{p},i\omega_p)=[Z(i\omega_p)+\omega_p\delta Z(i\omega_p)
/\delta \omega_p]\tau_3$, partly because no appropriate information on 
$\delta \Sigma(p)/\delta \mu$ is available and partly because we know that 
this difference, which appears only in a single term in the infinite sum of 
$T\sum_{\omega_{p'}}$, becomes very important only when the small-polaronic 
effect is large~\cite{Takada95c}.

%%%%%%%%%%%%%%%%%   Paragraph 23: Equations to determine Tc  %%%%%%%%%%%%%%%%%%%
By summarizing all the above assumptions and considerations, we can derive a 
couple of equations, one for $Z(i\omega_p)$ and the other for $\phi(i\omega_p)$, 
to determine $T_c$ by retaining only up to the linear order in $\phi(p)$ 
in Eq.~(\ref{eq:22}) in the following way:
\begin{align}
\label{eq:33}
Z(i\omega_p)\!=&\! 1\! + \!\frac{2T}{\omega_p} \!\sum_{\omega_{p'}}
\lambda(\omega_{p'}\!-\!\omega_p)\Gamma(i\omega_{p'},i\omega_p)
\nonumber \\
&\times \tan^{-1}\!\left [\!\frac{\omega_c}{\omega_{p'}Z(i\omega_{p'})}\!\right ],
\\
\Delta (i\omega_p) \!=&\! \frac{2T}{Z(\omega_p)}\!\sum_{\omega_{p'}}\!
\frac{\Delta (i\omega_{p'})}{\omega_{p'}}
[\lambda(\omega_{p'}\!-\!\omega_p)\!-\!\mu^*]
\nonumber \\
& \times \Gamma(i\omega_{p'},i\omega_p)
\tan^{-1\!}\left [\!\frac{\omega_c}{\omega_{p'}Z(i\omega_{p'})}\!\right ],
\label{eq:34}
\end{align} 
where $\Delta (i\omega_p)$ is defined by $\phi(i\omega_p)/Z(i\omega_p)$. 
We can define $T_c$ by the highest temperature at which nonzero 
$\Delta (i\omega_p)$ can be found. We shall call $T_c$ obtained from these 
equations a result in the GISC scheme. In the Eliashberg theory, $T_c$ is 
determined by the solution of Eqs.~(\ref{eq:33}) and (\ref{eq:34}) with 
taking $\Gamma(i\omega_{p'},i\omega_p)=1$. Usually, the Eliashberg function 
$\alpha^2 F(\Omega)$ is not renormalized in the process of self-consistently 
determining $Z(i\omega_p)$. In this sense, the Eliashberg theory might be better 
regarded as the ``GW$_0$'' approximation, rather than the GW one. If we do not 
solve Eq.~(\ref{eq:33}) for $Z(i\omega_p)$ but simply take $Z(i\omega_p)=1$ 
in the solution of Eq.~(\ref{eq:34}), the obtained $T_c$ may be considered 
as a result in the G$_0$W$_0$ approximation. 

%%%%%%%%%%%%%%%%%%%%%%%%%%%%%%%%%%< Figure 3 >%%%%%%%%%%%%%%%%%%%%%%%%%%%%%%%%%%
\begin{figure}[htbp]
\begin{center}
\includegraphics[scale=0.40,keepaspectratio]{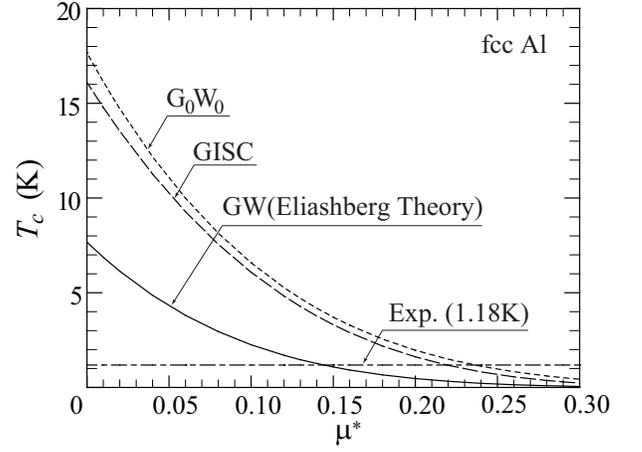}
\end{center}
\caption[Fig.3]{Superconducting transition temperature $T_c$ for fcc Al as a 
function of the Coulomb pseudopotential $\mu^*$ in three different calculation 
schemes in comparison with the experimantal $T_c$ of 1.18K.}
\label{fig:3}
\end{figure}
%------------------------------------------------------------------------------%

%%%%%%%%%%%%%%%%%%%%     Paragraph 24: Result for fcc Al   %%%%%%%%%%%%%%%%%%%%%
By using the $\alpha^2 F(\Omega)$ function in Fig.~\ref{fig:2}, we have 
calculated $T_c$ for fcc Al in three different schemes and shown the results 
as a function of $\mu^*$ in the Eliashberg theory (solid curve), the GISC 
(dashed curve), and the G$_0$W$_0$ (dotted curve) in Fig.~\ref{fig:3}. Note 
that this is a typical example for the weakly-correlated and weakly-coupled 
superconductors. The cutoff energy $\omega_c$ is increased up to 
$5\langle \Omega \rangle$ in order to obtain the convergent results for $T_c$. 
As can be seen in Fig.~\ref{fig:3}, $T_c$ in the most sophisticated GISC scheme 
is very different from that in the Eliashberg theory but very close to that 
in the simplest scheme of the G$_0$W$_0$, indicating that the vertex corrections 
included in the GISC is mostly cancelled by the self-energy renomarization 
corrections. 

%%%%%%%%%%%%%%%%%%%%%%%%%%%%%%%%%%< Figure 4 >%%%%%%%%%%%%%%%%%%%%%%%%%%%%%%%%%%
\begin{figure}[htbp]
\begin{center}
\includegraphics[scale=0.44,keepaspectratio]{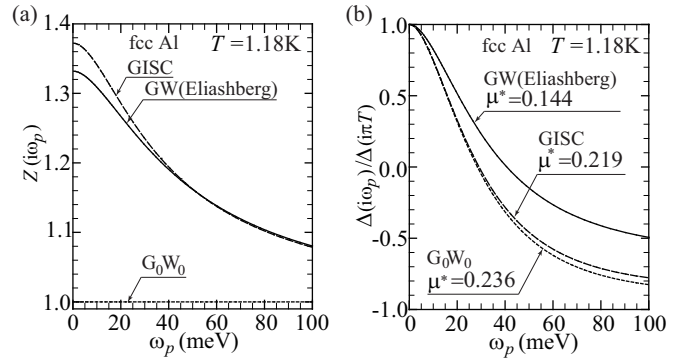}
\end{center}
\caption[Fig.4]{(a) Renormalization factor $Z(i\omega_p)$ and (b) the gap 
function $\Delta(i\omega_p)$ normalized by $\Delta(i\pi T)$ calculated for 
fcc Al at $T=1.18$K in the three different schemes.}
\label{fig:4}
\end{figure}
%------------------------------------------------------------------------------%

%%%%%%%%%%%%%%%%%%%%     Paragraph 25: Result for fcc Al   %%%%%%%%%%%%%%%%%%%%%
Because the experimental value of $T_c$ for fcc Al is 1.18K, we can determine 
the value of $\mu^*$ for each scheme; they are 0.144, 0.219, and 0.236 for 
the Eliashberg, the GISC, and the G$_0$W$_0$, respectively. Then, we can compare 
the results for $Z(i\omega_p)$ and $\Delta (i\omega_p)/\Delta (i\pi T)$ at $T=
1.18$K, as shown in Fig.~\ref{fig:4}. Although the Eliashberg and the GISC 
provide more or less the same $Z(i\omega_p)$, the results for 
$\Delta (i\omega_p)/\Delta (i\pi T)$ are very different. We see that as far as 
the superconducting properties are concerned, the G$_0$W$_0$ gives about the 
same quality of results as the GISC. 

%%%%%%%%%%%%%%%%%%%%%%     Paragraph 26: Discussion   %%%%%%%%%%%%%%%%%%%%%%%%%%
Four comments are in order: 
(i) According to the Migdal's theorem~\cite{Migdal58}, one might expect that 
the Eliashberg theory gives approximately the same $T_c$ as the GISC for such 
weakly-coupled conventinal superconductors because of the irrelevance of the 
vertex corrections. A closer inspection of the Migdal's proof reveals, 
however, that the theorem states only the irrelevance of the first-order vertex 
corrections, but not for the first-order self-energy corrections. If the 
irerelance of the former corrections is true, the same must be true for the 
latter, given that the WI always holds. In this regard, the Migdal's theorem 
actually proves the relevance of the G$_0$W$_0$ approximation, which is indeed 
confirmed by the results in Fig.~\ref{fig:3}.
(ii) The preference of the G$_0$W$_0$ approximation to the Eliashberg theory 
applies only to the calculation of $T_c$; the normal-state property in a metal 
as represented by $Z(i\omega_p)$ is better described by the latter, as can be 
seen in Fig.~\ref{fig:4}(a). 
(iii) Since $T_c$ in all the three schemes varies very much with the change of 
$\mu^*$, those schemes are not good enough for the first-principles calculation 
of $T_c$, as long as $\mu^*$ is given phenomenologically. Thus, we need to 
develop a scheme to determine $\mu^*$ from first principles. The present author 
pursued such a scheme in the framework of the G$_0$W$_0$ 
approximation~\cite{Takada78}, according to which the gap function is not 
considered as a function of the frequency variable but the momentum one and 
the gap equation to give $T_c$ is derived for the momentum-dependent 
gap function $\Delta({\bm p})$ as
\begin{align}
\Delta({\bm p})= -\sum_{\bm p'} \frac{\Delta({\bm p'})}{2\varepsilon_{\bm p'}} 
\tanh \left (\frac{\varepsilon_{\bm p'}}{2T_c} \right )
{\cal K}_{{\bm p},{\bm p'}},
\label{eq:F3}
\end{align}
where the pairing interaction ${\cal K}_{{\bm p},{\bm p'}}$ is defined by 
\begin{align}
{\cal K}_{{\bm p},{\bm p'}} \!=\! \int_0^{\infty} \!\frac{2}{\pi}
\,d\Omega\,\frac{|\varepsilon_{\bm p}|\!+\!|\varepsilon_{\bm p'}|}
{\Omega^2\!+\!(|\varepsilon_{\bm p}|\!+\!|\varepsilon_{\bm p'}|)^2}\,
W({\bm p}\!-\!{\bm p'},i\Omega),
\label{eq:F5}
\end{align}
with $W(q)$ given by Eq.~(\ref{eq:23}) in which $\Pi(q)$ is replaced by 
$\Pi_0(q)$. This scheme was successfully applied to real materials such as 
the $n$-type SrTiO$_3$~\cite{Takada80} and the graphite intercalation 
compounds~\cite{Takada82,Takada09a}.
(iv) First-principles determination of $\mu^*$ is also made possible by 
SCDFT~\cite{Oliveira88,Kurth99,Luders05,Marques05}, according to which the 
fundamental gap equation is just the same as that in Eq.~(\ref{eq:F3}), but 
the pairing interaction is different from the one in Eq.~(\ref{eq:F5}). 
Because the form in Eq.~(\ref{eq:F5}) includes the contribution from the 
plasmons~\cite{Takada78} in a very natural way~\cite{Arita13,Arita14}, 
it is strongly recommended to employ Eq.~(\ref{eq:F5}) for future 
implementation of SCDFT~\cite{Takada15}.
\medskip

%%%%%%%%%%%%%%%%%%%%%%%%%%%%%%%%%< Conclusion >%%%%%%%%%%%%%%%%%%%%%%%%%%%%%%%%
\section{Conclusion}

%%%%%%%%%%%%%%%%%%%%%%%%%%< Paragraph 27: Conclusion >%%%%%%%%%%%%%%%%%%%%%%%%%%
In summary, we have discussed the relevance of the G$_0$W$_0$ approximation 
in obtaining some of physical quantities in both normal and superconducting 
states. Due to the existence of the WI, this approximation remains 
to be useful in a much wider interaction range than one might naively imagine, 
probably up to the medium-coupling range in which most of the real materials 
are involved. 

\section*{Acknowledgements}

Y. Takada thanks H. Maebashi and S. Ishii for discussions in the normal state. 
This work is partially supported by Innovative Area 
"Materials Design through Computics: Complex Correlation and Non-Equilibrium 
Dynamics" (No. 22104011) from MEXT, Japan.

%%%%%%%%%%<  References  >%%%%%%%%%%

%%%%%%%%%%%%%%%%%%%%%%%%%%%%%%%%%%%%%%%%%%%%%%%%%%%%%%%%%%%%%%%%%%%%%%%%%%%%%%%%
\end{document}